# Software for Studying CASCADE Error Correction Protocols in Quantum Communications

Nikita Repnkiov
*Department of Complex Information Security of Computer Systems*
*Tomsk State University of Computer Systems and Radioelectronics*
Tomsk, Russia
repnikov-nik@mail.ru

Vladimir Faerman
*Department of Complex Information Security of Computer Systems*
*Tomsk State University of Computer Systems and Radioelectronics*
Tomsk, Russia
fva@fb.tusur.ru, 0000-0002-9643-0245

*Abstract*— This article addresses the development of quantum communication methods in the context of emerging quantum computing threats and emphasizes the importance of key reconciliation in quantum communication systems. The study focuses on the CASCADE protocol and the design of a software prototype intended for research and educational purposes. A parallel error-correction algorithm based on the actor model was implemented, improving the efficiency of key reconciliation and reducing the amount of exchanged data. Evaluation of the prototype revealed limitations, including the computational cost of message passing, complexity of error handling, and code redundancy due to iterative development. Experimental results confirmed the correct implementation of the core CASCADE algorithms and informed the design of future improvements. Proposed enhancements include redesigning the system architecture, developing interfaces for exporting intermediate data, defining the communication channel as a separate component, and expanding tools for systematic verification and comparative analysis of blind key-reconciliation methods.

*Keywords—CASCADE, QKD, Sequence Reconcilation, Error Correction, Actor Model, Syndrome.*

I. Introduction

An important component of information security is cryptographic protection. Its development has historically been shaped by the ongoing struggle between encryption methods and cryptanalysis. For example, early ciphers used before the 20th century were broken using frequency analysis. In particular, the classical Vigenère cipher, once considered completely secure with a sufficiently long key, eventually lost its resilience. Today, modern cryptographic systems use both symmetric and asymmetric methods, although the latter are not mathematically proven to be secure [1].

The advent of quantum computing in the coming decades threatens to compromise these systems, particularly key distribution schemes such as Diffie-Hellman and its variants. Two approaches are considered to address this challenge: increasing the complexity of mathematical methods (post-quantum cryptography) or using fundamental physical principles to generate identical random sequences on both sides of the exchange [2].

An identical random sequence can be used as a one-time pad (Vernam cipher) or to generate a symmetric key [2]. In both cases, it is essential to ensure that the key material remains identical despite potential errors in the communication channel. This process, known as key reconciliation, employs error-correction methods that do not reveal information, such as error-correcting codes. These methods rely on classical message exchange, which is not confidential, so protocols are designed to minimize information available to potential attackers [3]. As in other telecommunications applications, system performance is critical, influencing the requirements for both methods and protocols [4].

This work focuses on the CASCADE error-correction method [4]. The purpose of this article is to share experience in developing a prototype software suite for studying protocol variants. The suite is intended for research, group project-based learning, and advanced training in quantum communication.

II. Background

*A. Previous Study*

Within the research group, the post-processing of key material in the context of quantum key distribution (QKD), particularly raw sequence reconciliation, has been systematically studied. Reference [5] presents a decomposition of the key sequence post-processing task, while [3] compares CASCADE variants and classifies them based on their features, properties, and differences [6–9]. These studies provided the foundation for the CASCADE decomposition described in the qualification thesis.

Review studies [6, 10] indicate that the CASCADE protocol remains an important and actively studied solution in QKD, despite its main limitation: high interactivity. It serves as a benchmark for comparison with other methods, is widely implemented in commercial systems due to its efficiency and well-understood security, and provides a foundational model in educational applications.

*B. Previous Software Implementations*

Previous research on CASCADE variants involved multiple independent software implementations, each targeting specific tasks:

- **Python:** designed to study the overall protocol and its software primitives;
- **C++:** used to compare known CASCADE variants across metrics such as computation time.

Based on this experience, the current task was to create a software environment for systematic study of CASCADE variants and examination of their properties. This article presents the overall concept of the solution and describes the development and testing of the first prototype.

## III. DESIGN AND IMPLEMENTATION

### A. Design Goals

The proposed solution aims to create a flexible and intuitive environment that enables researchers without programming experience to study and simulate various CASCADE protocol implementations. The platform is designed to provide modularity, extensibility, and ease of organizing computational experiments, while maintaining transparency in the structure and interactions between entities. The main objectives include:

- creating a no-code platform for exploring CASCADE protocol modifications;
- implementing a mechanism for configurable, interchangeable components;
- logically separating key entities – Alice, Bob, Eve, and the classical channel;
- enabling simulation of any known CASCADE variant without modifying the architecture;
- providing tools to automate computational experiments and visualize results.

### B. Core Concept

The actor model is a computing paradigm in which the primary units are independent entities called actors. Each actor maintains its own internal state and a mailbox of messages, which it processes by creating new actors, updating its state, or sending messages to other actors. Unlike traditional shared-memory models, actors interact solely through message passing, offering isolation, modularity, and reliability in parallel computations [11].

The actor model was chosen for this platform due to the asynchronous nature of the CASCADE protocol, which runs in parallel on two machines with continuous message exchange. To accurately simulate this behavior, the architecture requires components that operate independently and interact asynchronously. The actor model fulfills these requirements, providing a straightforward mechanism for parallel event processing and state management without risking data races.

An additional benefit of this approach is the high scalability and integrability of the system [11]. Since actors do not share memory, standardization is needed only at the message level, which simplifies adding new components or modifying existing ones. Aligning actors with the physical entities of the protocol (Alice, Bob, Eve, and the classical channel) allows natural modeling of interactions and clear separation of accessible information. This design reflects real protocol conditions and enhances user understanding.

During prototype development, the actor architecture was implemented as specialized actors, each responsible for a specific system aspect:

- **Reconciliation Round Actor:** manages the logic of a single exchange round and parity checks;
- **Round Management Actor:** coordinates the sequence of rounds and manages transitions between them;
- **Intermediate Computation Display Actor:** visualizes the current state and process dynamics;
- **Computation Result Output Actor:** generates and presents the final experiment data.

Interactions between components are organized through strictly defined message formats, ensuring modularity, reproducibility, and scalability. The message exchange scheme and relationships between actors are shown in Fig. 1, 2, 3.

### C. Design Tools

The choice of programming paradigm and language was guided by requirements for reliability and formal correctness in modeling reconciliation routines. To meet these goals, a declarative functional paradigm was adopted, offering mathematical rigor and facilitating formal verification. This approach also complements the actor model, which is critical for asynchronous message exchange [12].

F# [13] was selected as the implementation language because of its integration with .NET, support for formal verification (via F⋆), and an eager evaluation model better suited for simulations than Haskell's lazy evaluation. It also provides concise syntax and benefits from an active community, making it preferable to Lisp, OCaml, and Standard ML [14]. Development was carried out in Visual Studio 2022 with a GitHub repository for version control and debugging.

### D. Core Algorithmic Operations

A key feature of the CASCADE protocol is that errors are detected and corrected based on information from previous rounds. In the initial prototype, bit masks were used to divide the distributed bit sequence. This method proved difficult to interpret and complicated debugging and analysis of the algorithm.

In the current implementation, a 4-color binary tree [15] represents the divisions, with nodes implemented as product types [16]. Nodes with previously calculated syndromes are shown in red, leaves corresponding to sequence elements with errors are blue, and compromised bits are yellow. The compiler represents these as nested structures, facilitating visualization and analysis.

Tree nodes and leaves are implemented as algebraic data types (tagged unions), consisting of two variants: leaf and tree node. The color of tree elements is also a sum type with two variants. The tree itself is represented as recursively nested structures, improving readability and simplifying region selection, while enabling potential optimizations [17]. Sample operations are shown in Fig. 4.

## IV. CASCADE BASIC SOFTWARE COMPONENTS

### A. Core Primitives

According to the previously performed decomposition of the CASCADE protocol, it can be reduced to a set of basic components. Depending on the choice and configuration of software primitives associated with these components, different protocol variants can be defined [6]. Within the prototype, the following variable components, which are integral to CASCADE, were considered:

- **Generation of round permutations:** used to shuffle the bit sequence when moving to a new round;
- **Division of the round sequence into blocks:** allows different options for selecting block size progression;
- **Break condition check:** determines when the algorithm should terminate.

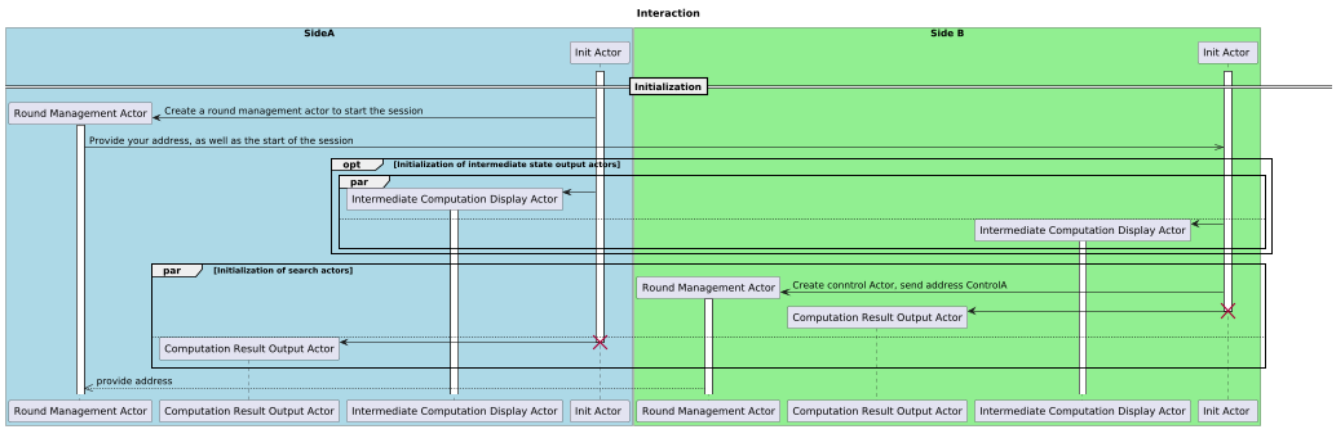

Fig. 1. UML sequence diagram for the protocol initialization process.
Blue represents the initiator (Alice), and green represents the responder (Bob).

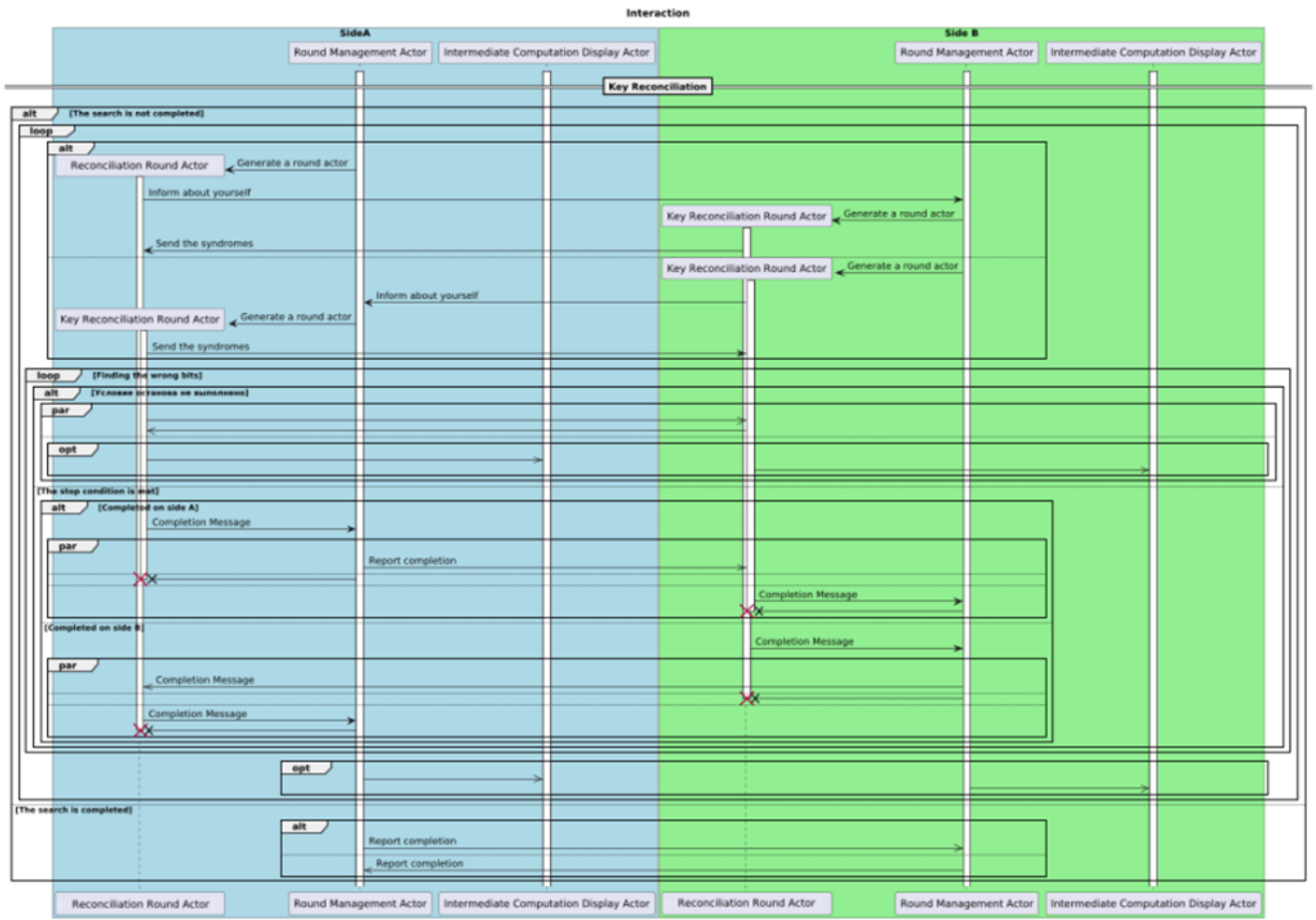

Fig. 2. UML sequence diagram for the distributed sequence reconciliation process.
Blue represents the initiator (Alice), and green represents the responder (Bob).

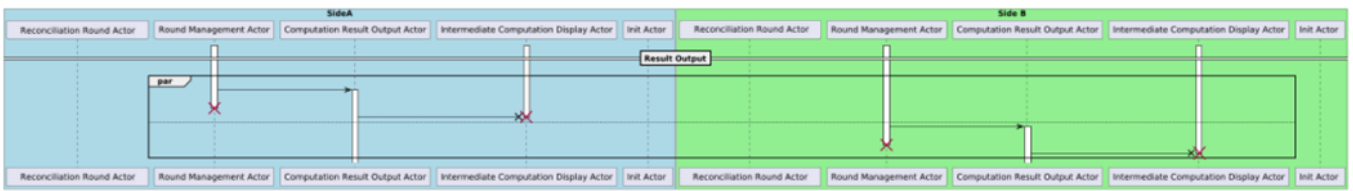

Fig. 3. UML sequence diagram for the protocol completion process.
Blue represents the initiator (Alice), and green represents the responder (Bob).

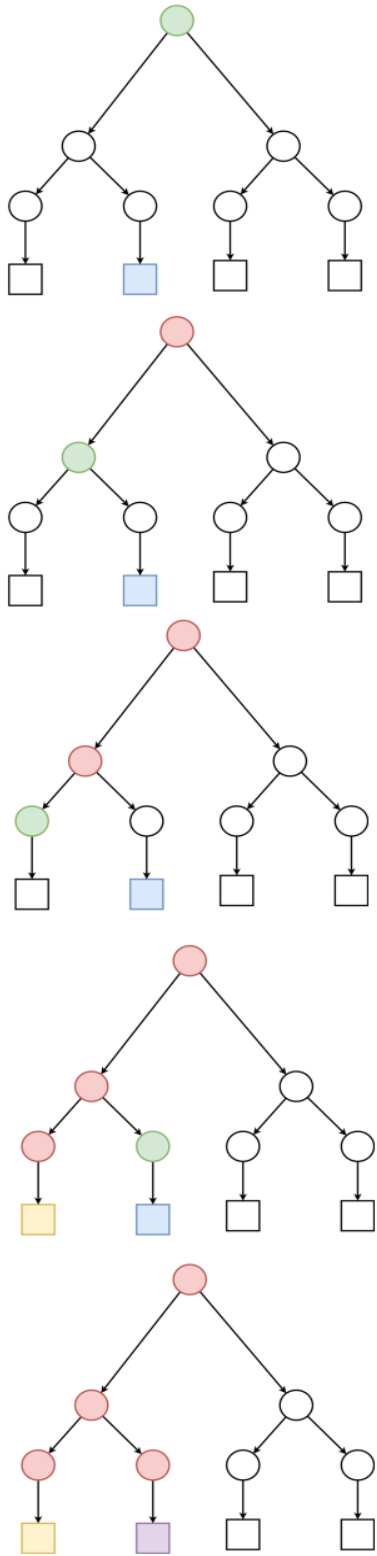

## B. Syndrome Aggregation

To reduce the interactivity of CASCADE, various methods of syndrome aggregation were explored to optimize the number of messages transmitted over the classical channel. As a result, a suboptimal variant was developed.

A combination of two methods was used: merging search trees and searching for multiple errors simultaneously. In the first method, tree nodes are only colored and do not lose their color, allowing new syndrome calculation data to be added without discarding previously obtained information. This enables combining results from different rounds and reduces the number of messages. An example is shown in Fig. 5.

Cascade search leverages a feature of the CASCADE protocol: detecting errors missed in previous iterations by analyzing corrected bits in new rounds. If the syndromes of two bit sequence segments matched in earlier rounds but changed after correction, this indicates an error in the corresponding segment. The search is performed in the second half of the smallest sequence containing the problematic bit, where a syndrome was previously calculated. The implementation is shown in Fig. 6, 7.

The algorithm generates a recursive list of leaves forming the path to the corrected bit and selects the first leaf for which the syndrome has not yet been calculated (shown in green). This localizes the search and reduces the number of messages, although in the presence of multiple errors, tree traversal is repeated.

To optimize multi-pass searching, a single-pass method was applied: the structure is wrapped in an *Option monad*, a new tree is created based on all paths to leaves with errors, the tree is reduced to elements where one subtree has a calculated syndrome and the other does not, and error search is performed on each subtree. The scheme is shown in Fig. 8, 9.

These methods significantly reduce the number of syndrome messages sent over the open channel without losing information about error locations, making the process more transparent and improving the performance of the CASCADE implementation.

Fig 4. Graphical representation of operations on a colored binary tree during bit error correction. Five consecutive transformations are shown from top to bottom.

The following algorithmic primitives were considered fixed rather than variable:

- **Bit error search algorithm (BINARY):** the core of CASCADE;
- **Parity reuse strategy:** involves storing the parities of subblocks from previous rounds.

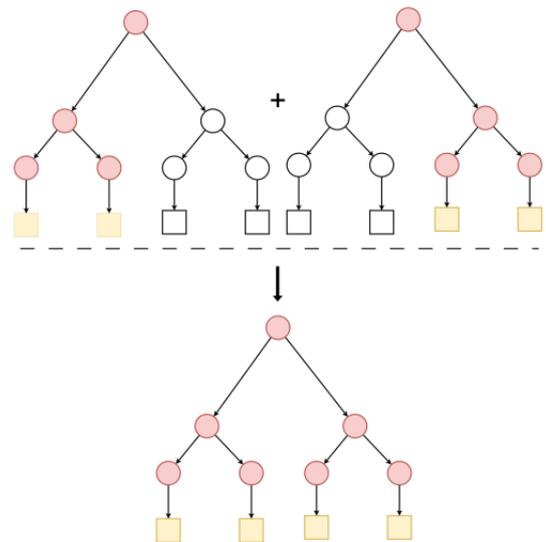

Fig 5. Graphical representation of the operation for merging two colored trees.

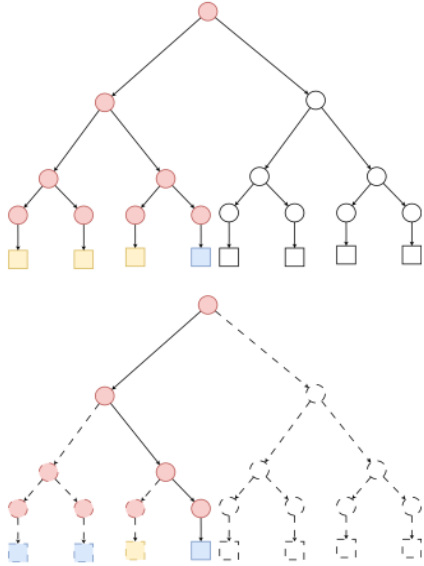

Fig 6. Graphical representation of the operation for locating an unvisited subtree.

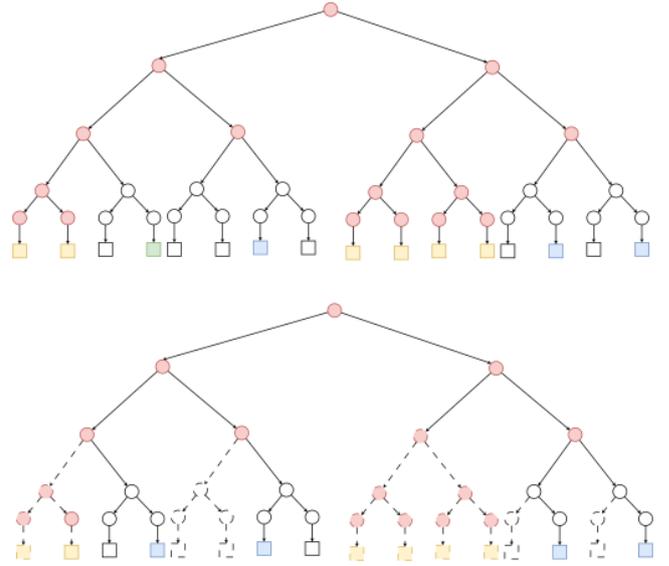

Fig 8. Graphical representation of the multi-path search of an unvisited subtree.

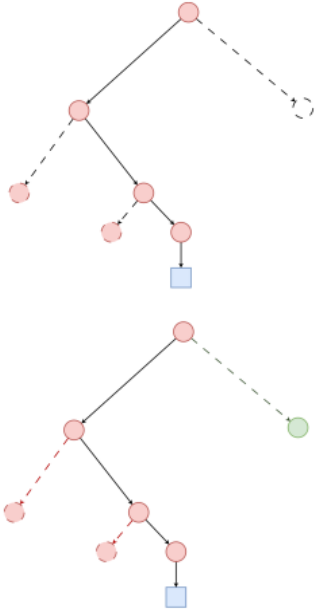

Fig 7. Graphical representation of the operation for locating an unvisited subtree in the remaining branch.

*2) Division of the round sequence into blocks* was performed either statically or dynamically. In the static approach, block sizes increased by a factor of $k$ in each round, starting from an initial size $N_0$. The initial block size was defined as $S_0 = 1 / QBER$, where $QBER$ is the known bit error rate. In the dynamic approach, the block size for each round was determined as $S_i = 1 / QBER_i$, where $QBER_i$ is estimated in each round based on the number of errors corrected in the previous round.

*3) Break condition check* was implemented using one of three approaches. The probabilistic approach required a predefined number of consecutive rounds with no errors corrected. The dynamic approach relied on comparing the number of corrected errors with a specified threshold. The static approach used a fixed, predetermined number of rounds.

## V. PERFORMANCE EVALUATION

### A. Implemented Components

Within the prototype, several alternative implementations were developed for each of the variable components described in Section 4. These alternatives were not intended for practical deployment and therefore were not based on CASCADE variants used in real QKD systems. Instead, simplicity was the primary design criterion, allowing straightforward debugging on short toy sequences.

*1) Generation of round permutations* was based on a deterministic shuffle algorithm applied sequentially several times. An alternative permutation generation method relied on sorting by complementary values produced using a linear congruential generator (LCG).

### B. Testing Routine

The testing aimed to verify the correctness of both the algorithmic solutions and the software implementation of the CASCADE module, as well as to confirm that identical bit sequences could be generated. It also assessed how the choice of algorithmic components affects reconciliation results, highlighting the solution's sensitivity to changes in the procedures used.

Testing was carried out under two main scenarios:

*1) Fixed sequence length with varying bit error rate.* The bit error probability ranged from 0.005, increasing in steps of 0.005, across an 60-step arithmetic progression.

*2) Fixed number of bit errors with increasing sequence length.* The sequence length varied from 512 bits to 20,480 bits in increments of 512 bits.

Each type of test was repeated three times, and no program failures were observed during the experiments. To evaluate sensitivity to component selection, the metric used was the number of parity bits transmitted over the channel, which reflects potential information leakage.

The testing methodology and presentation of results do not allow conclusions about the efficiency of the algorithms for several reasons:

- deliberately inefficient combinations of block size progression and break conditions were used;
- critical metrics, such as the probability of successful frame reconciliation, were not considered;
- test results combine data collected under different conditions, such as frames of varying lengths, due to the use of automated testing tools.

Despite these limitations, testing confirmed the correctness of the prototype's software implementation and demonstrated that different protocol components could be configured or varied. Based on the test results, graph is presented in Figs. 9. The effectiveness of measures to reduce the number of syndrome messages transmitted is in Fig. 10.

## VI. Lessons Learned

### A. Issues and Constraints

The main drawbacks of the developed algorithm and its implementation are:
- The technology stack is relatively uncommon, which may complicate maintenance by other developers.
- Message passing in F# actors, though lightweight compared to Erlang or Scala.Akka, remains costly in terms of CPU usage and memory.
- Error handling relies on a complex hierarchy of dedicated actors rather than simple exceptions or *Result monads*, necessitating a comprehensive error-control system for scaling.
- The architecture, including the actor hierarchy and message formats, was developed iteratively, leading to code redundancy and making debugging and further development more challenging.

### B. Notes and Observations

Analysis of experiments and the prototype revealed the following:
- The module correctly corrects bit sequences, confirming the proper implementation of the core CASCADE algorithms.
- Iterative development refined the algorithmic logic but introduced code redundancy; future work will require refactoring and careful architectural planning.
- Full comparison with known CASCADE variants was not performed because the prototype lacked intermediary actors for exporting intermediate data. Verification was limited to selected configurations compared with classic CASCADE and one modification using known subblock parities. Overall results were comparable, although systematic discrepancies were observed and will require future attention.

### C. Technical Insights

Based on the experience gained, the following recommendations are proposed for future work and application of the solution:
- Using the actor model positively influenced module decomposition, reducing development effort for future functional expansion.
- The original actor interaction scheme proved effective but requires interface redesign, as each functional extension increases boilerplate code. This can be addressed by reducing coupling: standardizing modules and configurations through abstract interfaces and aggregating them via a common mediator component.
- Treating the classical message channel as a separate entity is advisable, enabling future modeling of adversary intervention.
- While iterative development introduced code redundancy, it clarified priorities for the next prototype version: refining architecture and interfaces for exporting and processing intermediate data, implementing components to reproduce the main CASCADE variants, and performing systematic verification.

## VII. Conclusion

The implementation of the software module resulted in an error-correction algorithm for bit sequences based on the actor model. This enabled parallel error searching, reducing the overall reconciliation time across multiple sequences.

Measures were implemented to reduce the number of messages transmitted, including syndrome aggregation and partial reduction of compromised bits. For certain sequence variants, additional optimizations were applied to further minimize the amount of transmitted information.

The implemented software can be integrated into the .NET environment with support for native compilation, and it can also be realized in OCaml using a modular system similar to that of F#.

Building on the experience gained, the plan is to propose and justify a new architecture for the software suite, carry out comprehensive code refactoring, and expand the information support package for both the system and its individual components. Additionally, interfaces will be implemented to enable data export linked to the physical entities.

The set of components will be expanded to include sequence-reduction methods and additional techniques for adaptive block-size selection. The software suite is intended to serve as a tool for comparative analysis of blind reconciliation algorithms based on CASCADE and to explore an alternative 'blind' reconciliation approach incorporating components with machine learning methods.

## References

[1] Peter W. Shor. "Polynomial-time algorithms for primefactorization and discrete logarithms on a quantumcomputer". In: *SIAM review* 41.2 (1999), pp. 303–332.

[2] L. Chen et al. *Report on Post-Quantum Cryptography*.Tech. rep. National Institute of Standards and Technology, 2016. URL: https://nvlpubs.nist.gov/nistpubs/ir/ 2016/nist.ir.8105.pdf (accessed on 04/10/2025).

[3] E. V. Antropov et al. "Obzor protokolov ispravleniia oshibok Cascade i AYHI v sistemakh kvantovogo raspredeleniia kliuchei". Russian. In: *Vysokoproizvoditel'nye vychislitel'nye sistemy i tekhnologii* 8.2 (2024), pp. 42–57.

[4] C. H. Bennett, G. Brassard, and J.-M. Robert. "PrivacyAmplifcation by Public Discussion". In: *SIAM Journalon Computing* 17.2 (1988), pp. 210–229.

[5] E. I. Vasil'ev. "Etapy postobrabotki bitovykh posledovatel'nostei v ramkakh kvantovogo raspredeleniia kliuchei". Russian. In: *Sbornik izbrannykh statei nauchnoi sessii TUSUR*. Vol. 1-3. 2024, pp. 71–75.D.

[6] D. Elkouss, J. Martinez-Mateo, and V. Martin, "Information reconciliation for QKD," *Quantum Information & Computation*, vol. 11, no. 3–4, pp. 226–238, 2011, doi: 10.26421/QIC11.3-4-3

[7] Tupkary and N. Lutkenhaus. "Using Cascade in ¨quantum key distribution". In: *Physical Review Applied*20.6 (2023), p. 064040.

[8] G. Brassard and L. Salvail. "Secret-Key Reconciliationby Public Discussion". In: *Advances in Cryptology– EUROCRYPT '93*. Ed. by T. Helleseth. Vol. 765.Lecture Notes in Computer Science. Berlin, Heidelberg:Springer Berlin Heidelberg, 1994, pp. 410–423.

[9] Hao Yan et al. "Information Reconciliation Protocol in Quantum Key Distribution System". In: *2008 Fourth International Conference on Natural Computation*. IEEE, 2008, pp. 637–641. DOI: 10.1109/icnc.2008.755. URL: http://dx.doi.org/10.1109/ICNC.2008.755 (accessed on 04/10/2025).
[10] E. I. Vasil'ev and V. A. Faerman. "Diskretno sobytiinaia model' opticheskogo kanala sistemy kvantovogo raspredeleniia kliuchei". Russian. In: *Vysokoproizvoditel'nye vychislitel'nye sistemy i tekhnologii* 8.1 (2024), pp. 83–92.
[11] W. Clinger. "Foundations of Actor Semantics". PhD Thesis. USA: Massachusetts Institute of Technology, 1981.
[12] J. Backus. "Can programming be liberated from the von Neumann style? A functional style and its algebra of programs". In: *Communications of the ACM* 21.8 (1978). Turing Award Lecture, pp. 613–641.
[13] *The F# Software Foundation*. F# Software Foundation. URL: https://fsharp.org/ (visited on 04/10/2025).
[14] Kh. Abel'son and D. D. Sassman. *Struktura i Interpretatsiia Komp'iuternykh Programm*. Russian. 1st. 195 p. 1985.
[15] C. Okasaki. "Preface". In: *Purely Functional Data Structures*. Cambridge: Cambridge University Press, Apr. 1998, pp. ix–x.
[16] P. Martin-Lof. *Intuitionistic Type Theory*. Vol. 1. Studies in Proof Theory. Notes by Giovanni Sambin. Naples: Bibliopolis, 1984.
[17] Mads Tofte. "Type inference for polymorphic references". en. In: *Inf. Comput.* 89.1 (Nov. 1990), pp. 1–34.

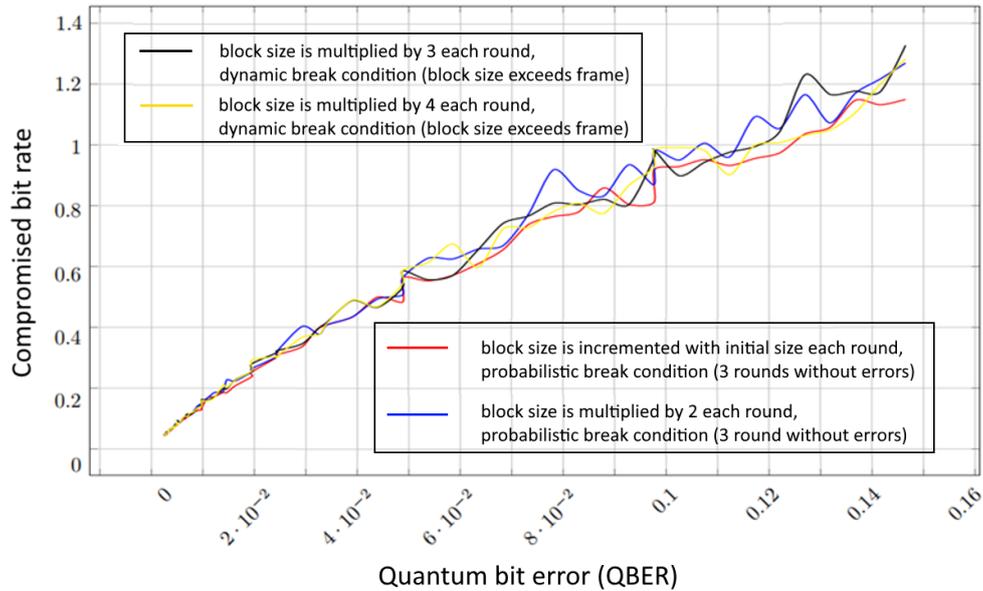

Fig. 9. Dependence of the fraction of compromised parity bits on the bit error probability in the quantum channel. The graphs were generated from tests with varying bit sequence length (from 512 bits to 20480 bits). Low correction efficiency is due to short reconciled frames and suboptimal algorithm parameter choices.

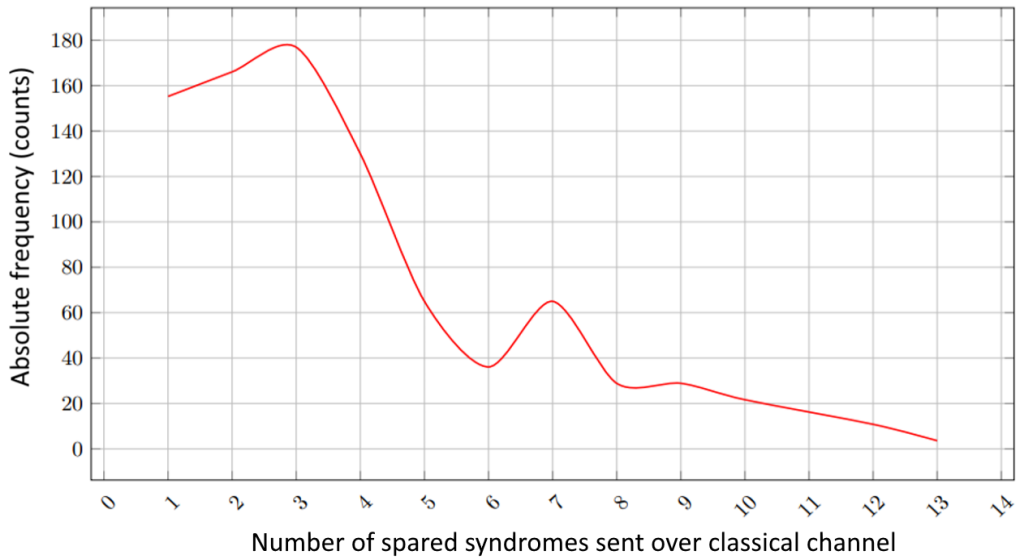

Fig. 10. Distribution of the reduction in the number of syndrome messages transmitted over the classical channel in the optimized algorithm version. The distribution was generated from tests with varying bit sequence length (from 512 bits to 20480 bits) and different bit error probabilities in the quantum channel. In the absence of a standardized implementation of classical CASCADE, the reduction in message count was evaluated relative to the prototype's initial implementation.